\documentclass[universe,article,accept,pdftex,moreauthors]{Definitions/mdpi} 

\usepackage{bm}
\usepackage{upgreek}

\newcommand{\apj}{Astrophys. J.}
\newcommand{\apjs}{Astrophys. J. Suppl. Ser.}
\newcommand{\apjl}{Astrophys. J.}
\newcommand{\mnras}{Mon. Not. R. Astron. Soc.}
\newcommand{\na}{New Astron.}
\newcommand{\aap}{Astron. Astrophys.}
\newcommand{\pasj}{Publ. Astron. Soc. Jpn.}

\newcommand{\Bpol}{B_{\rm pol}}

\firstpage{1} 
\pubvolume{1}
\issuenum{1}
\articlenumber{0}
\pubyear{2024}
\copyrightyear{2024}
\externaleditor{Academic Editor: Name}
\datereceived{2 April 2024} 
\daterevised{1 May 2024} 
\dateaccepted{6 May 2024} 
\datepublished{ } 
\hreflink{https://doi.org/} 

\Title{Hourglass Magnetic Field of a Protostellar System}

\TitleCitation{Hourglass Magnetic Field of a Protostellar System}

\Author{Shantanu Basu 
 $^{1,}$*\orcidA{}, Xiyuan Li $^{1}$\orcidB{} and Gianfranco Bino $^{2}$}

\AuthorNames{Shantanu Basu, Xiyuan Li and Gianfranco Bino}

\AuthorCitation{Basu, S.; Li, X.; Bino, G.}

\address{%
$^{1}$ \quad Department of Physics and Astronomy, University of Western Ontario, London, ON N6A 3K7, Canada; xli2522@uwo.ca 
\\
$^{2}$ \quad State Street Corporation, 
 30 Adelaide Street East \#1100, Toronto, ON M5C 3G6, Canada; gbino@uwo.ca
}

\corres{Correspondence: basu@uwo.ca} 

\abstract{
An hourglass-shaped magnetic field pattern arises naturally from the gravitational collapse of a star-forming gas cloud. Most studies have focused on the prestellar collapse phase, when the structure has a smooth and monotonic radial profile. However, most observations target dense clouds that already contain a central protostar, and possibly a circumstellar disk. We utilize an analytic treatment of the magnetic field along with insights gained from simulations to develop a more realistic magnetic field model for the protostellar phase. Key elements of the model are a strong radial magnetic field in the region of rapid collapse, an off-center peak in the magnetic field strength (a consequence of magnetic field dissipation in the circumstellar disk), and a strong toroidal field that is generated in the region of rapid collapse and outflow generation. A model with a highly pinched and twisted magnetic field pattern in the inner collapse zone facilitates the interpretation of magnetic field patterns observed in protostellar clouds.
}

\begin{document}




\section{Introduction}
\label{sec:intro}
The general picture of star formation involves the gravitational collapse of a dense subregion (a core) of a molecular cloud that is threaded by an ambient magnetic field. The near flux freezing at low densities
leads to an hourglass magnetic field configuration in a star-forming core. While predicted theoretically many decades ago \cite{mestel66,mouschovias76}, the observational evidence for hourglass fields emerged more recently \cite{schleuning98,girart06,girart09}. These observations infer a projected magnetic field through dust polarization \cite{girart09,qiu14,soa18,sad19} or dichroic extinction \cite{Kandori_2017}. They tend to confirm that 
gravitational collapse begins with a relatively well-ordered magnetic field and that the turbulent energy does not dominate the magnetic energy within collapsing cloud cores. However, many of these sources are protostellar, which are expected to have a complex three-dimensional magnetic field structure.

The hourglass pattern is a good model for the prestellar phase of collapse, before a central star--disk system has formed. During the prestellar phase, the magnetic field can be modeled as essentially poloidal because the rotation rate and strength of the toroidal magnetic field $B_\phi$ remain small, i.e., $B_\phi \ll B_r < B_z$ is the expected hierarchy. In the later protostellar phase, the rapid infall onto the protostar and the accompanying rotational spinup create significant values of $B_r$ and $B_\phi$ in the inner regions. The dissipation of the magnetic field by ohmic dissipation and ambipolar diffusion also cause a distortion of the magnetic field pattern. In this phase, $B_\phi \sim B_r \sim B_z$ in some inner regions. This transition can be seen in the simulations of \cite{machida08} and subsequent works. 

An analytic solution for a prestellar hourglass-shaped magnetic field was derived in cylindrical coordinates by \cite{ewertowski13}. They calculated the magnetic field directly from Ampere's Law upon assuming an axisymmetric electric current density. This solution was used successfully to fit the magnetic field in simulations of collapsing cores \cite{ewertowski13} and the magnetic field pattern inferred from polarized starlight in the prestellar core FeSt--457~\cite{binobasu2021}. In the latter case, ref.
~\cite{binobasu2021} found that the radial scale length $R$ of the magnetic field is significantly larger than the observed radial scale of $R_{\rm gas}$ of the gas distribution. This could be understood by the occurrence of significant flux redistribution by ambipolar diffusion in the formation phase of the core. See \cite{binobasu2021} for details. 

In order to evaluate the observed magnetic field of protostellar systems, it is helpful to build an analytic or semi-analytic model for this phase, when the radial and toroidal field components are significant or even dominant in some regions. Prior to this work, the only analytic magnetic field model applicable to the protostellar phase was the magnetized toroid model of \cite{li96}. That model applied to the pivotal moment when a protostar first forms, and does not include the effects of rotation and the disk and outflow that subsequently develop. Previous work to model dust polarization maps in the protostellar phase has utilized detailed simulation results \cite{tomisaka11, kataoka2012}, but analytic functions can simplify the analysis and reduce computation time. In this paper, we introduce an axisymmetric semi-analytic model for the protostellar phase that captures many features of simulation results that evolve well into the Class 0 phase and include a disk, outflow, and nonideal MHD \cite{basu2024}. These features include an off-center peak in $B_z$ and significant amplitudes of $B_r$ and $B_{\phi}$.


\section{Methods}
\label{sec:methods}

\subsection{Magnetic Field}

An axisymmetric poloidal magnetic field can be derived from Ampere's Law using cylindrical coordinates and by describing the electric current density $\bm{j} = j(r,z)\hat{\phi}$. The poloidal field 
 $\bm{B}_{\rm pol} = B_r\hat{r} + B_z\hat{z}$ can be obtained from the vector potential $\bm{A} = A(r,z)\hat{\phi}$. It was shown by \cite{ewertowski13} that, if the current density is separable, i.e.,
\begin{equation}
  j(r,z) = f(r)g(z),  
\end{equation}
and $g(z) = e^{-z^2/h^2}$, then, in the domain $r \leq R$,
\begin{equation}
\label{eq:A_series}
	A(r,z) =\sum_{m=1}^{\infty}  k_m J_1(\sqrt{\lambda_m}  r) \left[ \mathrm{erfc} \left( \frac{\sqrt{\lambda_m} h}{2} + \frac{z}{h} \right)e^{\sqrt{\lambda_m} z}  + \mathrm{erfc} \left( \frac{\sqrt{\lambda_m} h}{2} - \frac{z}{h} \right)e^{-\sqrt{\lambda_m} z} \right] \, ,
\end{equation}
where
\begin{equation}
    k_m = \frac{2\, h \pi^{3/2} e^{(h^2 \lambda_m)/4}}{c\,R^2 \sqrt{\lambda_m}\,\left[ J_2\,(\sqrt{\lambda_m} R)\right]^2} \int_{0}^{R} f(r')\,  J_1(\sqrt{\lambda_m}\, r^\prime)\, r^\prime\, d r^\prime \, , \quad m \in \mathbb{N}\, ,
\label{eq:km}
\end{equation}
and
\begin{equation}
  \lambda_m = \left(\frac{a_{m,1}}{R}\right)^2,  
\end{equation}
in which $a_{m,1}$ is the $m$th positive root of $J_1(x)$, and $a_{m,1} < a_{m+1,1}$. 
Here, $R$ is the outer radius of the solution, and the magnetic field is assumed to equal $\bm{B}_0 = B_0 \hat{z}$ for $r>R$. 
See \cite{ewertowski13} for a detailed derivation and description of the boundary conditions adopted in this solution. Given these results and the relation $\bm{B} = \nabla \times \bm{A}$, along with the existence of a background magnetic field $\bm{B}_0 = B_0 \hat{z}$, the axisymmetric poloidal magnetic field components are

\begin{equation}
 B_r =\sum_{m=1}^{\infty}  B_m\, J_1(\sqrt{\lambda_m}  r) \left[ \mathrm{erfc} \left( \frac{\sqrt{\lambda_m} h}{2} - \frac{z}{h} \right) e^{-\sqrt{\lambda_m} z}   - \mathrm{erfc} \left( \frac{\sqrt{\lambda_m} h}{2} + \frac{z}{h} \right)e^{\sqrt{\lambda_m} z} \right] \, , \\
 \label{eq:Br2}
 \end{equation}
   
\begin{equation}
B_z =\sum_{m=1}^{\infty}  B_m\, J_0(\sqrt{\lambda_m}  r) \left[ \mathrm{erfc} \left( \frac{\sqrt{\lambda_m} h}{2} + \frac{z}{h} \right)e^{\sqrt{\lambda_m} z}  \\ + \mathrm{erfc} \left( \frac{\sqrt{\lambda_m} h}{2} - \frac{z}{h} \right)e^{-\sqrt{\lambda_m} z} \right] + B_0\, ,\label{eq:Bz2}
\end{equation}
where $B_m \equiv k_m \sqrt{\lambda_m}$ and carries units of the magnetic field. 

Equation (\ref{eq:km}) shows that, for any given radial current distribution function $f(r)$, one can evaluate the coefficients $k_m$ that yield the poloidal magnetic field components. Note that $j(r,z)$ is the induced current due to self-inductance and (at least partial) flux freezing during the gravitational contraction of a cloud, whereas a current density on a much larger scale is responsible for the background magnetic field $B_0$.
In \cite{bino2022}, various forms of the radial current density function $f(r)$ were used to derive hourglass magnetic field patterns. The adopted forms of $f(r)$ were all centrally peaked, specifically a Gaussian, a Bessel function $J_0$, and a smoothed power law. These were mostly applicable to prestellar cores, for the reasons described earlier. Here, we consider more complex forms that capture the physics of the protostar--disk--outflow system. In \cite{basu2024}, it was shown through three-dimensional simulations that the early protostellar (Class 0) phase of evolution is characterized by an off-center peak in $B_z$. This is due to the nonideal MHD effects that become significant at the typical density of protostellar disks. Furthermore, there are significant amplitudes of $B_r$ in the inner collapse zone due to rapid infall with magnetic field dragging and of $B_\phi$ in the innermost regions of rapid rotation and strong outflow.

We let the radial current distribution function be defined by two peaks and a power law tail as follows:
\begin{equation}
 f(r) = A\cdot e^{-(r-r_0)^2/\sigma_0^2}+B\left[S_{10}(r) \cdot e^{-(r-r_1)^2/\sigma_1^2} + S_{01}(r) \cdot \frac{a_0^2}{a_0^2+(r-r_1)^2}\right],
    \label{eq:radial current density model}
\end{equation}
where $A$ and $B$ are scaling factors that carry units of current density, $S_{10} = [1+(r_1-r)/|r_1-r|]/2$, and $S_{01} = [1-(r_1-r)/|r_1-r|]/2$. The step functions $S_{10}$ and $S_{01}$ switch from 1 to 0 or vice versa as $r$ crosses $r_1$, ensuring that the vicinity of the second peak is a Gaussian for $r < r_1$ and a power law for $r > r_1$. Here, $r_0$ and $r_1$ are the locations of the two separate peaks. The first peak has a Gaussian standard deviation $\sigma_0$. The second peak has a Gaussian inner side with standard deviation $\sigma_1$ and an outer side that is a smoothed power law with scale length $a_0$ and asymptotic dependence $\propto r^{-2}$.

To produce an off-center magnetic field peak that reflects the numerical simulation results in \cite{basu2024}, the scaling factors $A$ and $B$ usually take negative and positive values, respectively. 
While a single centrally peaked Gaussian in the radial current density $f(r)$ yields a centrally peaked poloidal magnetic field \cite{bino2022}, a double-peaked $f(r)$ with $A<0$ and $B>0$ is capable of yielding an off-center peak in the poloidal magnetic field.
A value of $A < 0$ is physically representative of an inner region of magnetic field dissipation; the magnetic field due to the inner negative (counterclockwise direction) electric current partially offsets the magnetic field due to the outer positive (clockwise direction) electric current. The second peak necessarily has a $B > 0$, and the outer power law decrease in $f(r)$ is representative of an extended region of infall. 
The location $r_1$ of the second peak in $f(r)$ sets the approximate location of the peak poloidal field strength on the midplane of the model. We note that, while the function $f(r)$ is an ansatz, the solutions for the magnetic field satisfy Maxwell's equations such that they are self-consistent.

A nonzero toroidal magnetic field that has significant amplitude in the inner region and especially in the outflow zone is a special feature of the protostellar phase. 
In the earlier prestellar phase, $B_\phi$ is expected to be small, and one can fit a projected magnetic field from polarization maps with a purely poloidal magnetic field model \cite{binobasu2021}. In the protostellar phase, there is significant $B_\phi$ in the rapidly rotating disk region, as well as in the outflow zone, where the inertia of the outflow can twist the ambient magnetic field. The projection of the toroidal field can significantly affect a polarization map of a star-forming region~\cite{tomisaka11,kataoka2012}, so it is vital to include it in a model of the protostellar magnetic field. These can aid in the interpretation of the polarization maps of protostellar outflow systems \cite{qiu14,lee19}. 

To model the toroidal field $B_\phi$, we note that there is a physical correlation between $B_r$ and $B_\phi$ within the region of rapid collapse (i.e., inside the expansion wave front---see \cite{shu77}). The region of strongly pinched magnetic field is also a region of strong twist in the magnetic field due to inward collapse with at least partial angular momentum conservation. The components $B_\phi$ and $B_r$ are both antisymmetric across $z=0$. Hence, it is reasonable to expect that $B_\phi$ is directly proportional to $B_r$. However, simulations also show that the region of enhanced $B_\phi$ is larger than that of enhanced $B_r$ due to twisting of the ambient magnetic field by an outflow \cite{basu2024}. We achieve a suitable balance of accuracy and simplicity by introducing a multiplicative dependence on $B_z$ as well, and, preferentially in regions somewhat above and below the midplane $z=0$, by introducing the vertical scale height $H$. This leads to the empirical formula
\begin{equation}
   B_\phi = a_1\, B_r\, \frac{B_z}{B_0} \,\left[ 1 - e^{-z^2/H^2} \right] \, ,
\end{equation}
where $a_1 > 0$ is a fitting coefficient. 

Altogether, our model can be cast in a normalized form for maximum generality and ease of computation. 
We define normalized quantities $\tilde{r} = r/R, \; \tilde{z} = z/R, \; \eta = h/R$, as well as $\tilde{B}_r = B_r/B_0, \; \tilde{B}_z = B_z/B_0, \; \beta_m = B_m / B_0$. This leads to normalized equations
\begin{align}
\tilde{B}_r = \sum_{m=1}^{\infty} \beta_m \,J_1(a_{\text{m},1} \tilde{r})\left[ \text{erfc} \left( \frac{a_{\text{m},1} \eta}{2} - \frac{\tilde{z}}{\eta} \right)e^{-a_{\text{m},1} \tilde{z}}  \right.  \left. - \text{erfc} \left( \frac{a_{\text{m},1} \eta}{2} + \frac{\tilde{z}}{\eta} \right)e^{a_{\text{m},1} \tilde{z}} \right] ,
\label{dim:br}
\end{align}
\begin{align}
\tilde{B}_z = \sum_{m=1}^{\infty} \beta_m \,J_0(a_{\text{m},1} \tilde{r})\left[ \text{erfc} \left( \frac{a_{\text{m},1} \eta}{2} + \frac{\tilde{z}}{\eta} \right)e^{a_{\text{m},1} \tilde{z}}  \right.  \left. + \text{erfc} \left( \frac{a_{\text{m},1} \eta}{2} - \frac{\tilde{z}}{\eta} \right)e^{-a_{\text{m},1} \tilde{z}} \right] + 1 .
\label{dim:bz}
\end{align}

Furthermore,
\begin{equation}
    \tilde{B}_\phi = a_1\, \tilde{B}_r\, \tilde{B}_z \,\left[ 1 - e^{-\tilde{z}^2/\tilde{H}^2} \right] \, ,
\end{equation}
where $\tilde{H} = H/R$.
In addition to picking $B_0$ as the unit of magnetic field and $R$ as the unit of length, we choose $c$ as the unit of speed. Hence, the unit of current density is $B_0\, c/R$. Therefore, the normalized counterpart of Equation (\ref{eq:radial current density model}) is
\begin{equation}
 \tilde{f}(\tilde{r}) = \tilde{A}\cdot e^{-(\tilde{r}-\tilde{r}_0)^2/\tilde{\sigma}_0^2}+
 \tilde{B}\left[S_{10}(\tilde{r}) \cdot e^{-(\tilde{r}-\tilde{r}_1)^2/\tilde{\sigma}_1^2} + S_{01}(\tilde{r}) \cdot \frac{\tilde{a}_0^2}{\tilde{a}_0^2+(\tilde{r}-\tilde{r}_1)^2}\right],
    \label{eq:current density normalized}   
\end{equation}
where $\tilde{f}$, $\tilde{A}$, and $\tilde{B}$ are current densities normalized by $B_0\, c/R$, and all length variables with a ``tilde'' are normalized by $R$. Equation (\ref{eq:current density normalized}) is used to obtain normalized coefficients
\begin{equation}
       \beta_m = \frac{2 \, \eta \pi^{3/2} e^{(\eta^2 a_{m,1})/4}}{\left[ J_2\,(a_{m,1})\right]^2} \int_{0}^{1} \tilde{f}(\tilde{r}^\prime)\,  J_1(a_{m,1} \, \tilde{r}^\prime)\, \tilde{r}^\prime\, d \tilde{r}^\prime
\label{eq:km norm} 
\end{equation}
for use in Equations (\ref{dim:br}) and (\ref{dim:bz}).


\subsection{Polarization Model}

We use the three-dimensional radiative transfer code POLARIS \cite{reissl16} to produce synthetic polarization maps from our magnetic field model. POLARIS simulates the scattering, absorption, and emission processes of the dust grains within the domain of our model. Details of the radiative transfer technique, including the adopted dust emissivity, absorption, and scattering properties, can be found in \cite{reissl16}. We assume a dust-to-gas mass ratio of 0.01, and other adopted dust grain properties are listed in Table 1 of \cite{binobasu2021}. We assume perfect alignment of dust grains through the action of radiative anisotropic torques (RAT); therefore, the calculated polarization percentages may be taken to be upper limits. In this study, we focus on the spatial maps of the variations in polarization percentage rather than absolute values of the same.

\section{Results}

\subsection{Magnetic Field Model} \label{sec:Bfield}


We use the formulation described in Section~\ref{sec:methods} to find a model that is representative of the protostellar disk-outflow region in the early (Class 0) phase of star formation. In this phase, theoretical calculations \citep{basu2024} show that there is an off-center peak in the magnetic field strength approximately at the edge of the centrifugal disk. The peak magnetic field occurs at the interaction zone between the outwardly diffusing magnetic field of the disk and the inwardly advecting magnetic field of the pseudodisk.

\begin{figure}[H]
\includegraphics[width=13.5 cm]{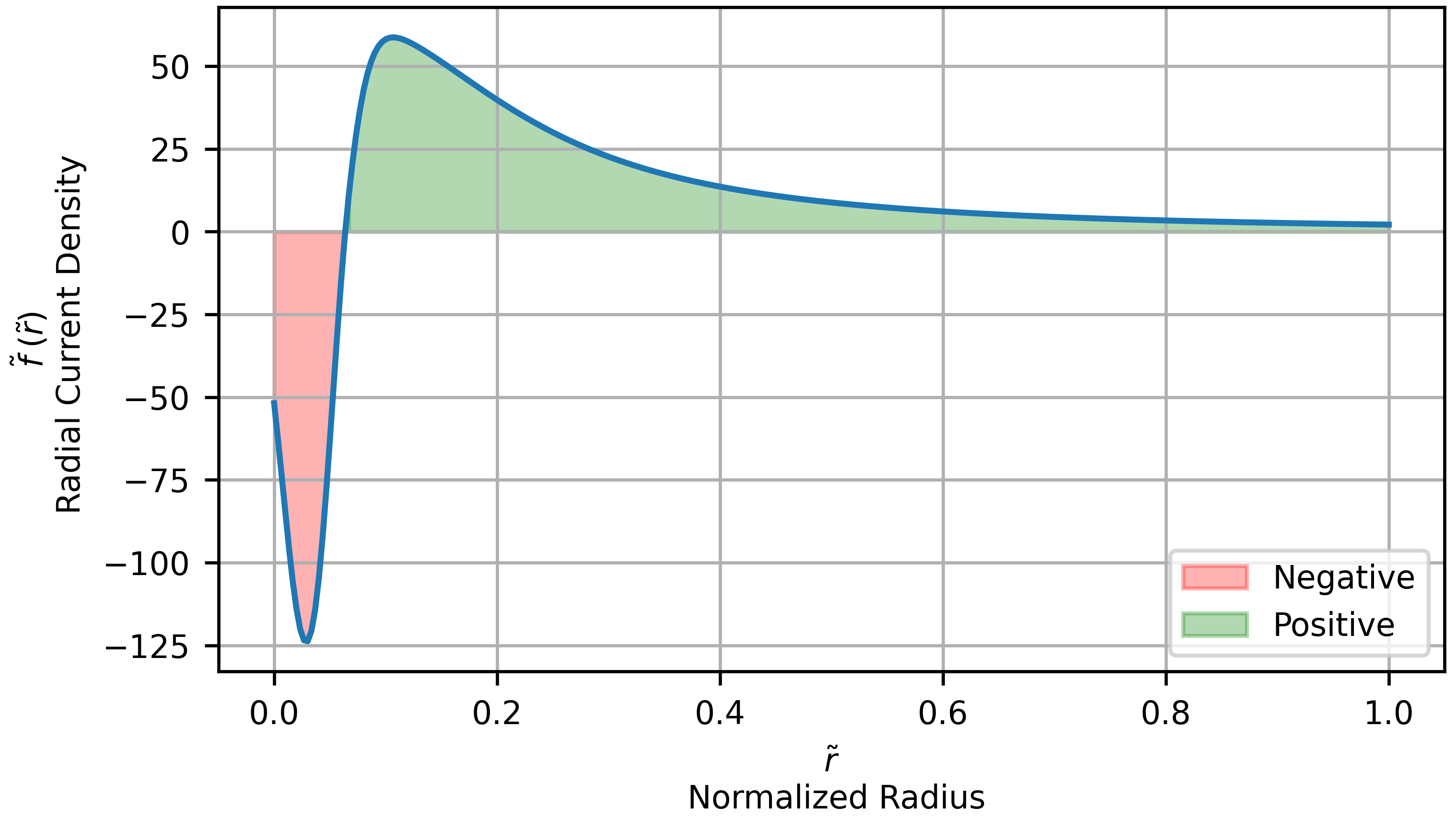}
\caption{The 
normalized radial current density $\tilde{f}(\tilde{r})$ versus normalized radius $\tilde{r}$ in our model. 
The function $\tilde{f}(\tilde{r})$ comes from Equation (\ref{eq:current density normalized}), and we use the parameter values listed in Table~\ref{tab:params}.}
\label{fig:current}
\end{figure}

\begin{figure}[H]

\begin{adjustwidth}{-\extralength}{0cm}
\centering 
\includegraphics[width=15 cm]{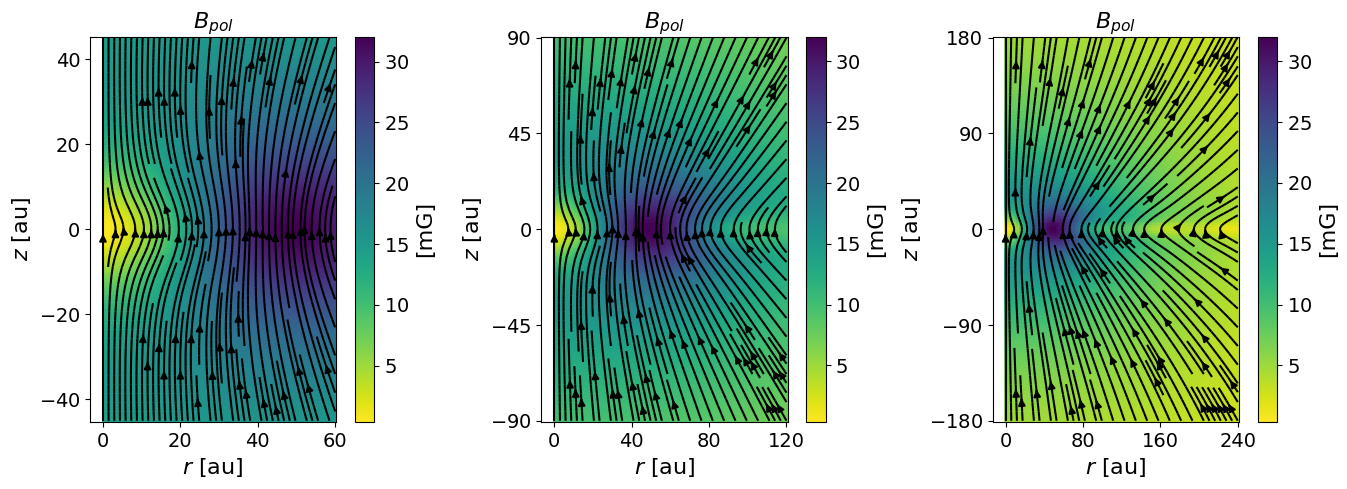}
\end{adjustwidth}
\caption{Lines 
 of poloidal magnetic field at three different scales, along with a color table of the field strength in units of mG.}
\label{fig:multiscale}
\end{figure}

Table~\ref{tab:params} lists the parameters used in our magnetic field model. Figure~\ref{fig:current} shows the radial dependence $\tilde{f}(\tilde{r})$ of the current density in our model, including a region of negative values at small radii that captures the effect of local magnetic field dissipation. Figure~\ref{fig:multiscale} shows the poloidal magnetic field structure and heat map on three different scales. Dimensional numbers are obtained with an adopted radius $R= 600$ au and background magnetic field $B_0 = 1.62$ mG. We choose $r_1 = \tilde{r}_1 R = 40$ au, which is physically associated with the disk edge. The remaining parameters in Table~\ref{tab:params} are chosen to give a good approximation of the magnetic field snapshot in Figures 3 and 10 of \cite{basu2024}
.
For the parameter values listed in Table~\ref{tab:params}, we find that keeping the first 17 terms in the series expansions of Equations (\ref{dim:br}) and (\ref{dim:bz}) gives accurate results. A much larger number of terms in the series is needed to be kept than in models of the prestellar phase \cite{binobasu2021} in order to resolve the small-scale structure of the off-axis magnetic field strength peak. Note that the model is meant to fit the region well within $r=R$ and not the outer collapsing core envelope that is at $r > R$. With the adopted parameters, the magnetic field profile in the inner few hundred au resembles the magnetic field profiles of the three-dimensional nonideal MHD simulation presented by \cite{basu2024}---see their Figure 7. 
The left panel of Figure~\ref{fig:multiscale} shows that the poloidal magnetic field strength has an off-center peak, which is physically associated with the disk edge. The middle panel shows the transition to a flared magnetic field of the surrounding pseudodisk at $r > 50$ au. The right panel emphasizes the flared field lines of the infalling pseudodisk. Thus, this model captures the inner region with an outwardly diffused magnetic field as well as the surrounding pseudodisk containing an inwardly advected magnetic field that is dragged in with near flux freezing.

\vspace{-9pt}
\begin{table}[H] 
\caption{Parameter values used in the magnetic field model. \label{tab:params}}
\newcolumntype{C}{>{\centering\arraybackslash}X}
\begin{tabularx}{\textwidth}{CCCCCCCCCC}
\toprule
\boldmath$\eta$ 
 & \boldmath$\tilde{A}$ & \boldmath$\tilde{B}$ & \boldmath$\tilde{r}_0$ & \boldmath$\tilde{\sigma}_0$  &  \boldmath$\tilde{r}_1$  & \boldmath$\tilde{\sigma}_1$   & \boldmath$\tilde{a}_0$ & \boldmath$a_1$ & \boldmath$\tilde{H}$  \\ 
\midrule
0.017		& $-$
144			& 63&  0.033 & 0.033 & 0.066 & 0.066 & 0.175 & 5 & 0.1 \\
\bottomrule
\end{tabularx}                                                                                              
\end{table}

 \vspace{-6pt}

\begin{figure}[H]

\begin{adjustwidth}{-\extralength}{0cm}
\centering 
\includegraphics[width=15 cm]{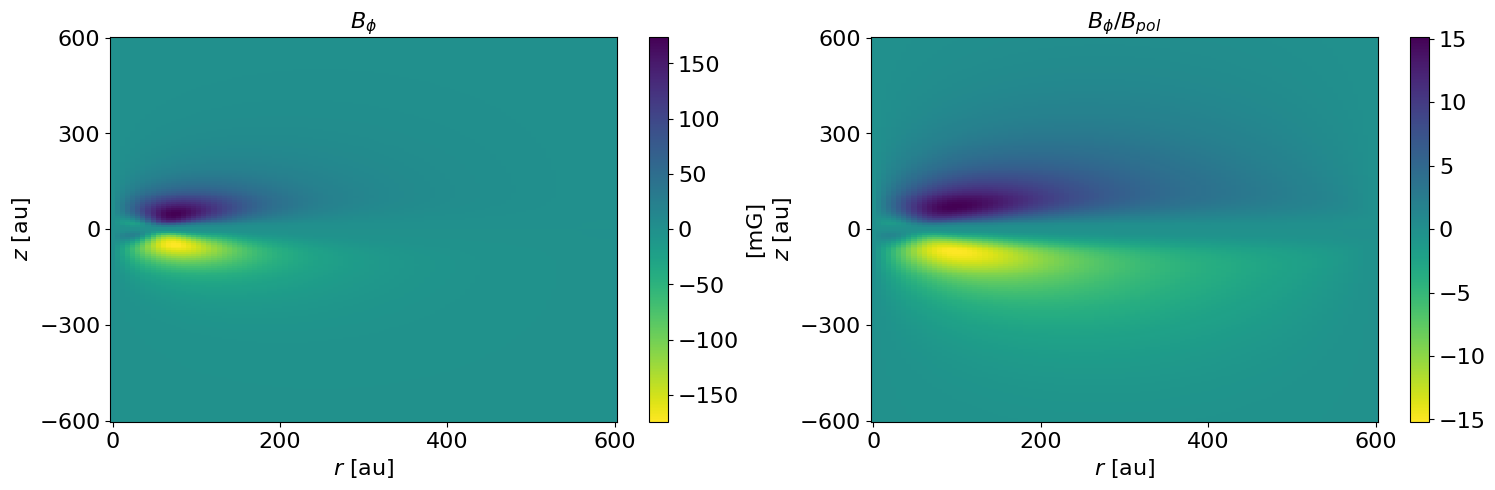}
\end{adjustwidth}
\caption{\textbf{Left
}: heat map of the toroidal magnetic field $B_\phi$. The color bar is in units of mG. \textbf{Right}: heat map of the ratio $B_\phi/\Bpol$.}
\label{fig:toroidal heat map}
\end{figure}

The full picture of the magnetic field, affected by rapid rotation and an outflow, is revealed by including the toroidal magnetic field $B_\phi$. 
During the protostellar phase, the rapid infall within an outgoing expansion wave will spin up the gas due to near angular momentum conservation. A wide-angle outflow launched primarily from the disk--pseudodisk interaction region \cite{basu2024} also transports angular momentum upward, which, in turn, generates significant magnitudes of $B_\phi$ above and below the midplane.
Figure~\ref{fig:toroidal heat map} shows heat maps of the toroidal field strength and the ratio $B_\phi/\Bpol$. These figures capture some key features of nonideal MHD simulations of the disk, pseudodisk, and outflow regions \cite{basu2024}---see their Figure~10. 
 This includes off-center peaks of $|B_\phi|$ above and below the midplane and an increasing ratio $|B_\phi/\Bpol|$ along the approximate direction of the outflow that emerges from just outside the disk.

 \vspace{-6pt}

\begin{figure}[H]
\includegraphics[width=11.5 cm]{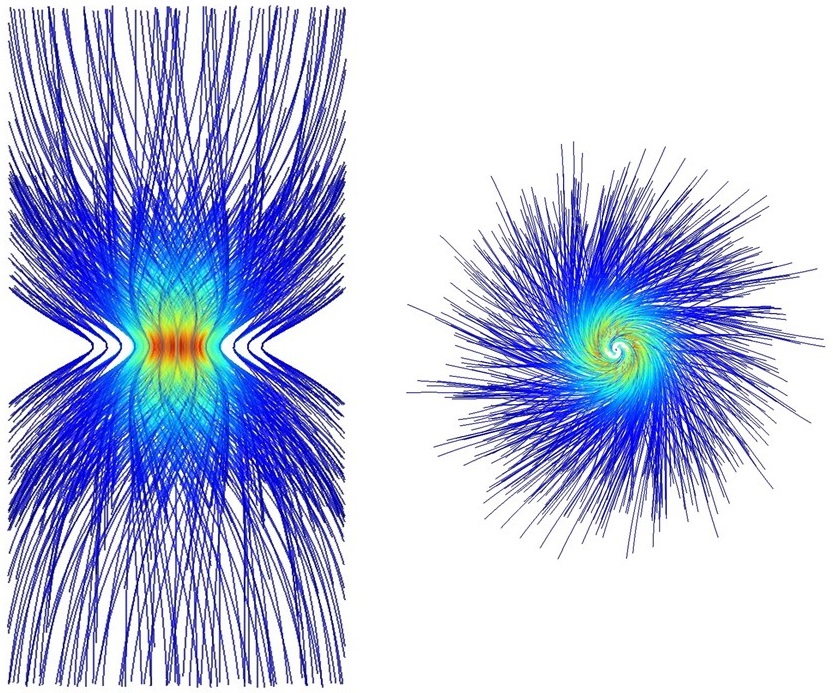}
\caption{\textls[-15]{Magnetic field streamlines. 
\textbf{Left}: a view from $90^\circ$ relative to the magnetic axis. \textbf{Right}:~a~view from $0^\circ$ relative to the magnetic axis.
The colors represent the relative field strength, with red the strongest and dark blue the weakest. The streamlines are plotted using the software package MAYAVI~\cite{ram11}.}}
\label{fig:stream1}
\end{figure}

\begin{figure}[H]
\includegraphics[width=11.5 cm]{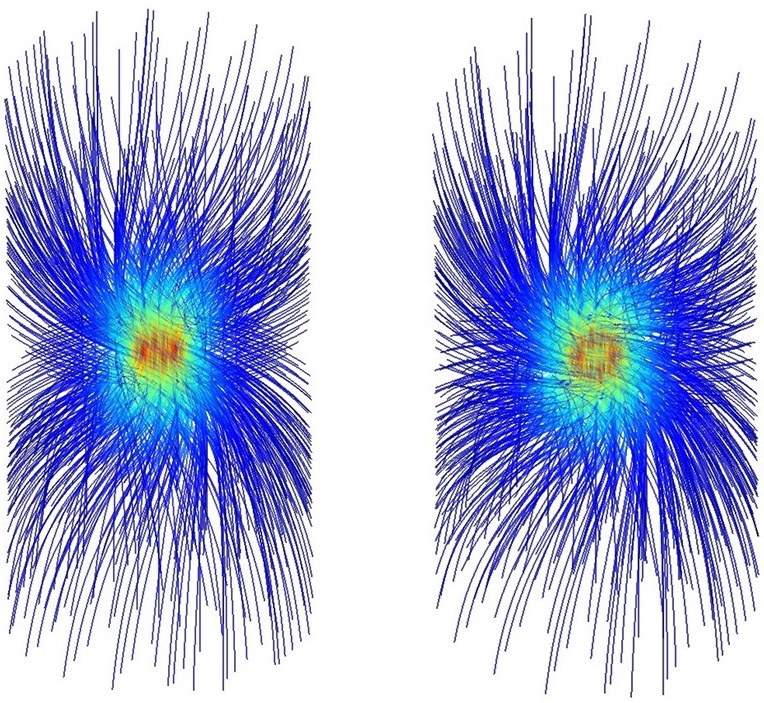}
\caption{\textls[-15]{Magnetic field streamlines. The plots are made in the same manner as in Figure~\ref{fig:stream1}.
\textbf{Left}: a~view from $60^\circ$ relative to the magnetic axis. \textbf{Right}:~a~view from $45^\circ$ relative to the magnetic axis.
The colors represent the relative field strength, with red the strongest and dark blue the weakest.} 
}
\label{fig:stream2}
\end{figure}

A visualization of the morphology of the total magnetic field is possible with a three-dimensional visualization of the streamlines of the magnetic vectors. {Figure~\ref{fig:stream1} shows the magnetic field streamlines of our model when viewed either perpendicular or parallel to the magnetic axis, respectively. Figure~\ref{fig:stream2} shows the corresponding views from angles tilted at $60^\circ$ and $45^\circ$ relative to the magnetic axis.
}  

\vspace{-9pt}

\subsection{Polarization Map} \label{sec:polmap}

To make further progress in understanding the possible observational realization of our magnetic field model, we generate a synthetic polarization map using the POLARIS radiative transfer code \cite{reissl16}. The dust density can be obtained from a gas density profile assuming a dust-to-gas mass ratio of 0.01.

To perform this calculation, we adopt an analytic gas density function that fits the main features of the inner collapse zone and an outflow cavity:
\begin{equation}
    \rho(r_s) = \rho_0 \left( \frac{r_s}{r_0}         \right)^\beta \, \sin^n \theta \, .
\end{equation}

This equation is written in the spherical coordinate system $(r_s,\theta,\phi)$ and is axisymmetric ($\phi$ independent).
Transformation to a cylindrical coordinate system is accomplished through $r_s = \sqrt{r^2 + z^2}$, $\theta = \tan^{-1} (r/z)$.
We adopt $\beta = -3/2$, the typical power law index for gravitational collapse inside an outgoing expansion wave \cite{shu77}. The factor $\sin^n \theta$ accounts for the cloud flattening and outflow cavity, similar to in models of magnetized toroids \cite{li96}, and the index $n$ can be adapted to fit the outflow zone calculated in simulations. Here, we adopt $n=2$. 

\begin{figure}[H]
\includegraphics[width=13.5 cm]{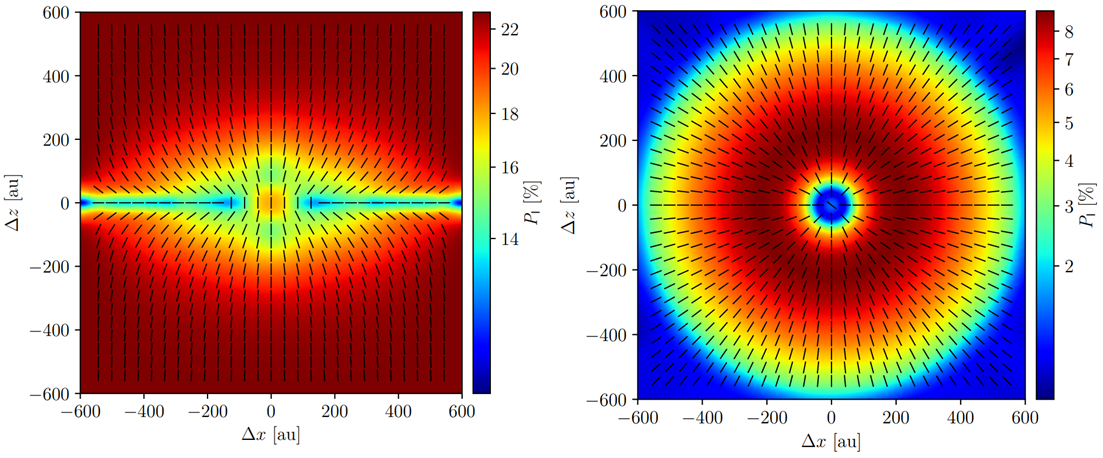}
\caption{Synthetic polarization maps of the poloidal magnetic field model. The vectors show the implied direction of the magnetic field (rotated by $90^\circ$ from the electric field polarization) and the polarization percentage in a color heat map. \textbf{Left}: a view from $90^\circ$ relative to the magnetic axis. \textbf{Right}: a view from $0^\circ$ relative to the magnetic axis.}
\label{fig:polaris_pol}
\end{figure}

Figure~\ref{fig:polaris_pol} shows segments of implied magnetic field direction (perpendicular to the emergent polarization direction) and a color map of the polarization percentage for the model with poloidal components only. We adopt $R=600$ au and $B_0$ = 1.62 mG as in previous plots. The two limits of viewing perpendicular to the magnetic axis (edge-on) and along the magnetic axis (pole-on) are shown. These figures reveal the well-known hourglass pattern in the edge-on plot and purely radial segments in the pole-on plot, for which the background field is along the line of sight. 
We focus here on the morphology of the polarization segments and the relative strength of the polarization rather than on the degree of polarization.


\vspace{-4pt}

\begin{figure}[H]
\includegraphics[width=13.5 cm]{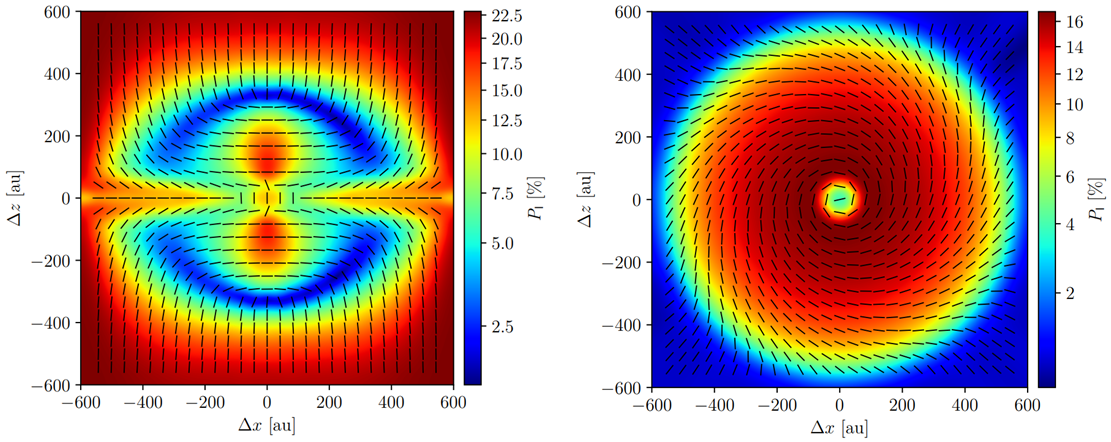}
\caption{Synthetic polarization maps of the total (poloidal and toroidal) magnetic field model. The vectors and heat map have the same meaning as in Figure~\ref{fig:polaris_pol}. \textbf{Left}: a view from $90^\circ$ relative to the magnetic axis. \textbf{Right}: a view from $0^\circ$ relative to the magnetic axis.}
\label{fig:polaris_tor}
\end{figure}

Figure~\ref{fig:polaris_tor} is the equivalent of Figure~\ref{fig:polaris_pol}, which shows the polarization map of the full magnetic field, including both poloidal and toroidal components. The edge-on map shows a clear transition from an hourglass at outer radii to purely horizontal segments at inner radii, where the toroidal field dominates. There is also a depolarization at some heights above and below the midplane where there is a significant volume of poloidal field in the foreground that is oriented at a large angle relative to the toroidal field in the inner region---see \cite{kataoka2012} for more discussion of this effect.
The pole-on map now emphasizes the circular pattern emerging from the toroidal field.

\vspace{-4pt}

\begin{figure}[H]
\includegraphics[width=13.5cm]{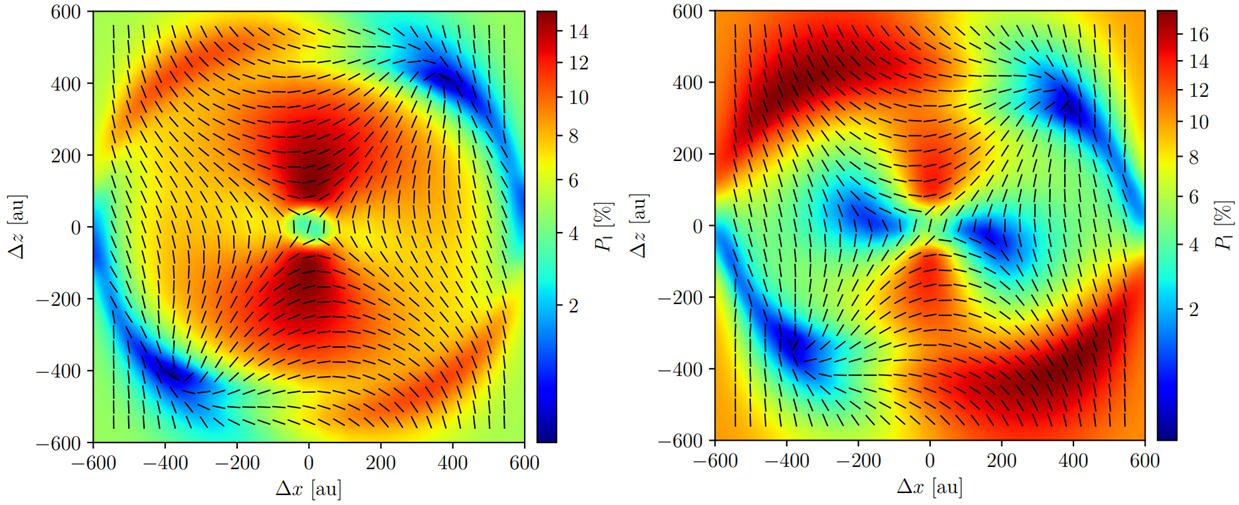}
\caption{Same as Figure~\ref{fig:polaris_tor} but with different viewing angles. \textbf{Left}: a view from $60^\circ$ relative to the magnetic axis. \textbf{Right}: a view from $45^\circ$ relative to the magnetic axis.}
\label{fig:polaris_tor2}
\end{figure}

Figure~\ref{fig:polaris_tor2} shows the polarization maps of the full magnetic field from two additional viewing angles. From these intermediate viewing angles, the implied magnetic field directions and distribution of polarized intensity are far more complex than in Figure~\ref{fig:polaris_tor}. An important feature is the pair of local polarization minima located at the top right and bottom left. There is also a pair of local polarization maxima at the top left and bottom right. The polarization map is not axisymmetric and also does not have a mirror symmetry about the midplane of the observation. This phenomenon was first demonstrated by \cite{tomisaka11}, who explained it in terms of the 
relative angles of $B_z$ and $B_\phi$ in the plane of sky when seen from different angles. These plots, in fact, have a point symmetry (symmetry under rotation by $180^\circ$).
This arises because $B_z$ is symmetric about $z=0$ while $B_\phi$ is antisymmetric about $z=0$. Observational evidence of such point symmetry in a polarization map would be strong evidence for the existence of significant toroidal magnetic fields \cite{tomisaka11,kataoka2012,lee19}. 
These plots also show a complex morphology that could be interpreted in terms of magnetic field line ``streamers'', i.e., radially directed magnetic field lines that point toward a central ``hub''. See \cite{beu20} for an observational example of such a hub--filament system in a region of massive star formation. Qualitative features of such a pattern can, in principle, emerge from a tilted view of a highly twisted but axisymmetric magnetic field.
\vspace{-4pt}
\section{Discussion}

We have demonstrated the ability of an analytic model to capture essential features that emerge from complex three-dimensional nonideal MHD simulations. These include an off-center peak in all three magnetic field components, a significant radial magnetic field near the pseudodisk, and an increasing relative strength of the toroidal field in regions associated with an outflow. A strength of the model is that the field components follow analytic expressions, with adjustable parameters that can be tuned to capture the extent of magnetic dissipation (through a negative current density at small radii near the midplane) and the amount of twisting of the magnetic field in the outflow zone.
Here, we have modeled the inner few hundred au squared region as the outflow zone around the protostar since this is where the greatest values of $|B_\phi|$ occur \cite{basu2024}. However, the full extent of outflowing gas can reach to at least thousands of au. Future work can explore models with a much larger outer radius $R$ than we adopted here. Such models will need to capture a larger dynamical range of scales and require a larger number of terms in the series solution of Equations (\ref{dim:br}) and (\ref{dim:bz}) than we have adopted. 

The analytic forms of the magnetic field and density profile ease the generation of synthetic polarization maps. The POLARIS code can take analytic functions as input, and runs much faster than when working with grid-based data. Analytic functions can also be tuned easily to reflect objects of different sizes and evolutionary stages.

The synthetic polarization maps show the complexity of interpreting observations. The toroidal field makes the projected magnetic field transition from an hourglass at outer radii to a series of horizontal segments at smaller radii. Therefore, the inferred magnetic field direction in these regions is perpendicular to the outflow direction. Observations that show the inferred magnetic field in perpendicular or random directions relative to the outflow direction \cite{hull19} have been used to suggest that the magnetic field is not dynamically important on scales smaller than $\sim 1000$ au. However, the dominance of the toroidal field at small radii means that there is a preference for the projected field to be perpendicular to the outflow when observed from $90^\circ$ relative to the magnetic axis. Furthermore, when viewing from an intermediate angle, the relationship is complex, and unusual patterns emerge.

The toroidal field used in our model is an empirical one that captures some of the essential features seen in simulations. Future work can refine the approach or start from the first principles of the current density, as performed for the poloidal field. The model has a peak in $B_\phi$ that is off-center in the radial direction, is not on the midplane $z=0$, and has a region of increasing $B_\phi/\Bpol$ where an outflow is expected to exist. These features are advantageous compared to a previous implementation of $B_\phi$ in the literature for fitting observational data~\cite{padovani13} where $B_\phi$ was chosen to be curl free and force free. The model of \cite{padovani13} results in $B_\phi$ having an $r^{-1}$ dependence, which does not have an off-axis peak and does not match the corresponding profiles emerging from simulations. We also expect that $B_\phi$ is not force free in that it should exert forces that drive the outflow.

Future work can leverage the simplicity of using the analytic forms of the magnetic field and density to make useful comparisons with observed polarization maps. Many of these maps are quite complex but may be probing systems that still have an underlying~symmetry.



\section{Conclusions}

We introduced an analytic magnetic field model for the protostellar phase. It captures essential physical features that define a protostellar disk--pseudodisk--outflow system. The model agrees well with snapshots of the Class 0 stage as calculated by \cite{basu2024}, including an off-center peak in the field strength and strong radial and toroidal field components. The analytic model facilitates the creation of synthetic polarization maps that can be compared with observations. The cases presented here show that observed polarization maps can appear quite complex, leading to difficulty in interpretation. This is the case even though the underlying three-dimensional pattern is quite ordered and contains a significant toroidal field component.






\vspace{6pt} 




\authorcontributions{Conceptualization, S.B.; methodology, S.B. and X.L.; software, X.L. and G.B; validation, S.B., X.L., and G.B.; formal analysis, S.B, X.L, and G.B.; writing---original draft preparation, S.B.; writing---review and editing, S.B. and X.L.; visualization, X.L. and G.B.; supervision, S.B.; funding acquisition, S.B. All authors have read and agreed to the published version of the~manuscript.}

\funding{This 
 work was supported by a Discovery Grant from the Natural Sciences and Engineering Research Council of Canada.}

\dataavailability{The data generated in the current study are available from the corresponding author on reasonable request.}

\acknowledgments{We thank 
 the referees for comments that improved the manuscript and Kohji Tomisaka for sharing valuable insights.}

\conflictsofinterest{Author Bino, G. is employed by the company State Street Corporation.
The authors declare that the research was conducted in the absence of any commercial or financial relationships that could be construed as a potential conflict of interest.
}

\abbreviations{Abbreviations}{
The following abbreviations are used in this manuscript:\\

\noindent 
\begin{tabular}{@{}ll}
MHD & Magnetohydrodynamics\\
POLARIS & Polarized Radiation Simulator
\end{tabular}
}

\begin{adjustwidth}{-\extralength}{0cm}

\reftitle{References}

\PublishersNote{}
\end{adjustwidth}
\end{document}